\documentclass[showpacs,showkeys,floatfix]{revtex4}
\usepackage[dvips]{graphicx}
\usepackage{epsfig}

\usepackage[cp1251]{inputenc}
\usepackage[english, russian]{babel}
\usepackage{color}
\usepackage{ulem}
\usepackage{bm}

\usepackage{feyn}

\begin{document}

\title{Why the smectic A -- hexatic phase transition does not follow its universality class?}

\author{E.I.Kats$^{1}$, V.V.Lebedev$^{1}$, and A.R.Muratov$^2$}

\affiliation{$^1$Landau Institute for Theoretical Physics, RAS, \\
142432, Chernogolovka, Moscow region, Russia \\
$^2$ Institute for Oil and Gas Research, Moscow, Russia.}

\begin{abstract}

We resolve the old riddle related to the critical behavior of the heat capacity near the smectic A -- hexatic second order  phase transition. Experiment suggests a ``large'' specific heat critical exponent $\alpha=0.5 \div 0.7$ inconsistent with the universality class for this phase transition implying the very small negative exponent $\alpha\approx-0.01$. We show that essential features of the heat capacity for the smectic A -- hexatic phase transition can be rationalized in the framework of a theoretical model treating jointly fluctuations of the hexatic orientational order and of the positional (translational) order parameters. Assuming that the positional (translational) correlation length $\xi_{tr}$ is larger than the hexatic correlation length $\xi_h$, we calculate a temperature dependence of the specific heat in the critical region near the smectic A -- hexatic phase transition. Our results are in a quantitative agreement with the calorimetric experimental data.

\end{abstract}


\maketitle

\section{Introduction}

Smectic liquid crystals are remarkable layered phases possessing an astonishingly rich variety of structures. The simplest smectic structure is the smectic A liquid crystal, which is solid-like in the direction perpendicular to the layers and fluid-like within the layers. By other words, the smectic A can be thought as a stack of liquid layers. Smectics C possess nematic-like orientationally ordered layers. In the late seventies, first as a theoretical suggestion \cite{BL78,BN81}, and later on by experimental observations \cite{PM81,BA86,JO03,ZK15} some smectic phases have been identified as stacks of layers with hexagonal orientational order, the phases are termed as hexatic smectics or hexatics. The hexatics possess a long-range hexagonal orientational order, however, they have no long-range positional (translational) order. It is worth to note that the hexatic phases are not merely a funny state of a few liquid-crystalline materials. They appear in many biological systems (see, e.g., Refs. \cite{NT00,MR05}), and even in planetary or astrophysical science as a form of dust plasma \cite{PV15}. Investigation of the hexatics is a multidisciplinary area including many fundamental physical problems and involving various questions of chemistry and biology. However, even after some decades of investigations of the hexatics, a complete description of this liquid crystalline state is still not available, and a number of phenomena remain to be clarified.

In this work we examine properties of a liquid crystal near the continuous smectic A -- hexatic phase transition, such transition is observed in a variety of materials, see Refs. \cite{PM81,BA86,ZK15,VL83,HV83,PN85,GS91,SG92,GS93,HK97,RD05,MP13}. We investigate the phase transition in spirit of the Landau approach in terms of the hexatic orientational order parameter. The order parameter for the phase transition has to be introduced from symmetry reasoning \cite{CL00}. Say, the nematic order parameter is a second order symmetric traceless tensor $Q_{ik}$ \cite{GP93}. In the nematic phase the average value of $Q_{ik}$ is nonzero and therefore the complete rotational symmetry of the isotropic liquid is reduced to the quadrupole symmetry of the nematic. An analogous ``nematic-like'' orientational order parameter can be introduced for the smectics $C$. It is also a second order symmetric traceless tensor, however, its components are within the smectic layers (i.e., the order parameter is a $2 \times 2$ tensor unlike the $3 \times 3$ order parameter in nematics). A non-zero average value of the order parameter means reduction of the uniaxial rotational symmetry $D_{\infty h}$, characteristic of the smectic layers in the phase A, to the biaxial rotational symmetry $D_{2h}$ in the phase C. By other words, in the smectics C the layers are invariant under rotation around the second order axis whereas in the smectics A the layers are isotropic.

Similarly to the smectic A -- smectic C phase transition the smectic A -- hexatic phase transition means a reduction of the rotational symmetry of the smectic layers. However, the symmetry of the smectic layers in the hexatics is $D_{6h}$ instead of $D_{2h}$ in the smectics C. By other words, in the hexatics the layers are invariant under rotations around the axis of the sixth order. The corresponding order parameter is a six-order symmetric irreducible tensor $Q_{injklm}$ (its irreducibility means $Q_{iijklm}=0$), having components solely within the layers. In the smectic A phase the average value of the tensor $Q_{injklm}$ is zero (the layers are isotropic) whereas in the hexatic phase its average value becomes non-zero. As a result, the smectic layers possess a hexagonal symmetry, i.e., they are invariant under rotation around the six-order rotation axis. The tensor $Q_{injklm}$ has two independent components \cite{GKL91}. Apart of small fluctuations, almost parallel smectic layers are well-defined. If  we direct the $Z$-axis perpendicular to the layers, then the two independent components of the tensor are $Q_{xxxxxx}$ and $Q_{xxxxxy}$. It is convenient to deal with the scalar complex field
 \begin{equation}
 \Psi =Q_{xxxxxx}+iQ_{xxxxxy},
 \label{complex}
 \end{equation}
instead of the tensor $Q_{injklm}$. At rotation by an angle $\chi$ around the $Z$-axis the order parameter $\Psi$ is transformed as
 \begin{equation}
 \Psi \to \exp(6i\chi) \Psi .
 \label{rotate}
 \end{equation}
The order parameter $\Psi$ is equivalent to the traditional  hexatic order parameter introduced in \cite{BL78,BN81} in terms of the molecular bond orientations (see also the textbooks \cite{CL00,GP93}).

The next step in constructing our theory is deriving the Landau functional for the smectic A -- hexatic  phase transition. Due to the rotational invariance the Landau functional ${\cal F}_{La}$ (determining an energy excess associated with the order parameter $\Psi$) contains only even in $\Psi $ terms in its expansion. Indeed, just the combinations $\Psi  \Psi ^\ast$, $(\Psi  \Psi ^\ast)^2$ etc are invariant under the transformation (\ref{rotate}). Therefore the smectic A -- hexatic phase transition should be of the second order (see the textbooks \cite{LL80,CL00,WK74,ST87}), in accordance with experiment. Since we deal with the two-component order parameter, the phase transition belongs to the same universality class as the superfluid one. Particularly, the heat capacity exponent $\alpha$ for the smectic $A$ -- hexatic phase transition should be small and negative $\alpha \approx - 0.01$ (see results of Monte Carlo simulations and experimental data presented in \cite{ST87,AN91,PV02}). Note to the point that the second order $\epsilon$-expansion \cite{GZ80} gives a positive but also very small (near $0.01$) value for the exponent $\alpha$. In a dramatic contradiction to this theoretical expectation, all known calorimetric data for the smectic A -- hexatic phase transition \cite{VL83,HV83,PN85,GS91,SG92,GS93,HK97,RD05,MP13} show large and positive exponent $\alpha = 0.5\div 0.7$. Thus one encounters an obvious problem, and the main motivation of this work is to find where is a catch.

We are not the first trying to resolve the contradiction. Some attempts have been performed in order to provide a rational basis for such behavior of the heat capacity. Two different suggestions can be found in literature. Having in mind that the exponent $\alpha$ is near $0.5$, it is tempting to assume that the smectic A -- hexatic phase transition occurs in a vicinity of a tricritical point \cite{AB86}. Then one could expect a sort of crossover behavior, in spirit of of the nematic -- smectic A -- smectic C tricritical point \cite{AV85}. However, it is hard to believe that all known hexatic liquid-crystalline materials, irrespective to the width of the stability region for the hexatic phase (ranging from a few to 50 degrees \cite{GS93}) are always near the tricritical point (and in the continuous phase transition side). Another suggestion as far as we aware advocated first in the works \cite{RH82,BB89,GC94} is based on an observation that measured experimentally critical exponents for the hexatic -- smectic $A$ phase transition are close to those predicted by the $q$-state Potts model \cite{BK76} (with $q=3$ or $q=4$ in 2- or 3-dimensional space). However it is not clear at all, how (and why) the Potts model can be mapped to the physics of the phase transition with the two-component order parameter. In our opinion, this approach is not consistent to be confronted with the entire array of the experimental data.

We propose another way to reconcile the well established phase transition universality conception and the massive of experimental data for the smectic A -- hexatic phase transition. A recent progress in experimental X-ray scattering techniques \cite{ZK15} reveals a rather unusual feature of the hexatics. Namely, very narrow peaks are observed in the X-ray scattering data there in the hexatic phase, that is the hexatics are almost ``ready'' to crystallize. The correlation length $\xi_{tr}$ of the short-range density fluctuations $\delta\rho$ (that is the positional order parameter), extracted from the X-ray data \cite{ZK15} for 3(10)OBC, is anomalously large, ranging from about $3\, nm$ in the vicinity of the transition point up to $20\, nm$ deeply in the hexatic phase. Then because of a relevant coupling between the hexatic and positional order parameters the universal critical behavior, characteristic of the superfluid helium universality class, should be observed only at the condition $\xi_h\gg \xi_{tr}$, where $\xi_h$ is the correlation length of the hexatic order parameter. In the opposite limit, at $\xi_h \leq \xi_{tr}$, the self-interaction of the hexatic order parameter mediated via the translational order parameter fluctuations is effectively non-local and, consequently, some non-standard critical behavior could be expected. It is the intermediate behavior which results in difficulties when developing theory tool. We examine just this possibility.

Of course, at the condition $\xi_h \gg \xi_{tr}$ one has to recover the standard universal behavior. However, to achieve such relatively large value of the orientational correlation length $\xi _h$ one has to probe a very narrow vicinity of the phase transition point. It would be fairly to say that from the available experimental data we are not in the position to estimate how narrow is this region of temperatures with the standard universal behavior. The textbook wisdom about second order phase transitions \cite{LL80,CL00} suggests only that $\xi _h \simeq \xi _0 |(T-T_h)/T_h|^{-\nu }$ where $\xi _0$ is the bare (microscopic) correlation length, $T_h$ is the second order smectic A -- hexatic transition temperature, and $\nu = 0.76$ is the critical exponent of the correlation length for the universality class with the two-component order parameter (in three-dimensional space). Assuming $\xi _0 \simeq 0.1\, nm$ we find that $\xi _h$ achieves the value a few times larger than $\xi_{tr} \simeq 3\, nm $ in the vicinity of the transition point  $\Delta T/T_h \lesssim 10^{-3}$. Beyond this region, one has to deal with the intermediate asymptotic behavior caused by the interaction of the hexatic order parameter $\Psi$ with the long-correlated translational order parameter fluctuations $\delta\rho$. The asymptotic behavior is characterized by a scaling behavior, and the main goal of our work is to construct a theoretical scheme for this case and to compare the theoretical predictions with known calorimetric experimental data.

Our paper is divided into the following sections. In the foregoing section \ref{sec:mod} we introduce ingredients and basic notions of our model and present its theoretical treatment. In the section \ref{sec:fluct} we examine fluctuation effects in the framework of our model. The section \ref{sec:exp} is devoted to confronting the theoretical results and experimental data. The section \ref{sec:con} summarizes our main findings.

\section{Model}
\label{sec:mod}

Our model is based on involving into consideration two strongly fluctuating fields: the hexatic order parameter $\Psi$ and the short-range density modulation $\delta\rho$ (that plays a role of the order parameter for the crystallization phase transition). We consider a vicinity of the smectic A -- hexatic phase transition where fluctuations of $\Psi$ are strong and one would expect the standard critical behavior with  universal exponents characteristic of a two-component order parameter. However, as we noted, the observed behavior of the heat capacity near the smectic A -- hexatic phase transition does not follow its universality class. We explain this unusual behavior by an interaction of the order parameter $\Psi$ with fluctuations of the short-range density modulation $\delta\rho$. Note to the point that the interaction is relevant also in the hexatic phase (out of the critical region), where effects related to the non-zero mean value of the order parameter $\Psi$ have to be included into the consideration. We defer the investigation of the region for a future work.

The cornerstone assumptions of our model are based on the experimental X-ray scattering patterns. It is well known, that the X-ray scattering is produced by the short-range electron density fluctuations proportional to the mass density fluctuations $\delta\rho$. Quantitatively, it is determined by the pair correlation function of $\delta\rho$. Fourier transform of the correlation function is known as the structure factor $S(\bm q)$:
 \begin{equation}
 S(\bm q)= \int d^3r\ \exp(-i \bm q \bm r)
 \langle \delta\rho(\bm r_1)
 \delta\rho(\bm r_1+\bm r) \rangle.
 \label{structure}
 \end{equation}
For the smectic phases, the structure factor $S(\bm q)$ has quasi-Bragg peaks at $q_x,q_y=0,q_z=2\pi/l$ (where $l$ is the thickness of the smectic layers) and also maxima at $q_\perp=q_0$, $q_\perp^2=q_x^2+q_y^2$, where $q_0^{-1}$ is on the order of a characteristic inter-molecular distance in the smectic layers. A hallmark feature of the hexatics, established experimentally \cite{ZK15}, is that near the cylinder $q_\perp=q_0$ the structure factor $S(\bm q)$ is almost independent of $q_z$ (more precisely, one should deal with the quasi-momentum and with the first Brillouin zone in $Z$-direction). In terms of the positional order parameter $\delta\rho$, the feature (weak dependence of the structure factor on $q_z$) implies that relevant fluctuations of $\delta \rho$ are strongly confined in the central part of the smectic layers.

The X-ray scattering data provide an information on the temperature dependence of the structure function $S({\bm q})$ near the smectic A -- hexatic phase transition (see \cite{ZK15} and references therein). In the smectic A phase the pattern is a bright diffuse ring (cylinder in the three-dimensional reciprocal space) parallel to the smectic layers with the radius $q_\perp=q_0$. In the hexatic phase the ring is split into six spots, according to the rotational six order  symmetry axis of the phase. The angular dependence of the structure function $S({\bm q})$ is related to a non-zero average value of the order parameter $\Psi$ in the hexatic phase. Each spot is narrow in the radial direction (along $q_\perp$) and elongated in the angular direction. The radial behavior of the structure function is characterized by the correlation length $\xi_{tr}$. As we already noted, for 3(10)OBC the correlation length $\xi_{tr}$ is ranging from about $3\, nm $ in the vicinity of the transition point up to $20\, nm$ in the hexatic phase. Any case, the correlation length is much larger, than the characteristic molecular size.

We analyze the critical behavior of a liquid crystal near the smectic A -- hexatic phase transition where fluctuations of the hexatic order parameter $\Psi$ are strong. In this case the angular dependence of the structure function in the hexatic phase is weak and therefore it will be neglected below. We are interested in the temperature region where the condition
 \begin{equation}
\xi _{tr} > \xi _h \ ,
 \label{corrl}
 \end{equation}
is valid (as above, $\xi _h$ is the hexatic order parameter correlation length, that characterizes correlations of the $\Psi$ fluctuations). As we explained in Introduction, the inequality (\ref{corrl}) is violated only in a very narrow vicinity of the smectic $A$ - hexatic transition temperature where the critical behavior characteristic of the superfluid phase transition has to be restored. Beyond the narrow vicinity, that is in the region where the inequality (\ref{corrl}) holds, a special theoretical analysis of the critical behavior is needed, that is the subject of our work.

Crystallization of a conventional liquid or of a liquid crystal is as a rule a strong first order phase transition. Particularly, its latent heat per molecule is on the order of (or larger than) $k_B T_m$, where $k_B$ is the Boltzmann constant, and $T_m$ is the crystallization temperature (room temperature in our case). Therefore one naturally expects the positional correlation length $\xi_{tr}$ to be on the order of the molecular scale $q_0^{-1}$. Since the experimental data \cite{ZK15} suggests $\xi_{tr}  \gg q_0^{-1}$, we do believe to the weak first order crystallization phase transition for the hexatics, and therefore the weak crystallization theory, see Ref. \cite{KLM93}, is a natural tool to describe positional fluctuations both, in the hexatic and in the smectic A phases, near the phase transition. Within this theory the characteristic value of the short scale density mass $\delta\rho$ (positional order parameter) is much smaller than the average uniform mass density. Then, according to the Landau approach, the energy, associated with fluctuations of $\delta\rho$ can be expanded into a series over $\delta\rho$. The expansion, known as the Landau functional, determines correlation functions of $\delta\rho$, particularly, the structure factor  (\ref{structure}).

The Landau functional contains some terms of the expansion in $\delta\rho$, starting from the second order contribution, which can be written as
 \begin{eqnarray}
 {\cal F}_{(2)}=\int \frac{d^3q}{(2\pi)^3}
 \left[ \frac{a}{2} |\delta\rho(\bm q)|^2
 +\frac{b}{2} (q_\perp-q_0)^2
  |\delta\rho(\bm q)|^2\right].
  \label{freeweak2}
 \end{eqnarray}
Here $\delta\rho(\bm q)$ is Fourier transform of $\delta\rho(\bm r)$ and $a$, $b$ are some coefficients. The coefficient $a$ diminishes as the temperature decreases, as usually in the Landau approach, we assume $a\propto(T-T_\star)$ where $T_\star$ is the bare crystallization temperature, i.e., the mean field stability limit of the hexatic state. The coefficient at $|\delta\rho(\bm q)|^2$ in Eq. (\ref{freeweak2}) has a minimum at $q_\perp=q_0$. It corresponds to the maximum of the structure function (\ref{structure}) at $q_\perp=q_0$. There is no dependence on $q_z$ in the expression (\ref{freeweak2}) in accordance with the discussed above experimental observations \cite{ZK15} showing no dependence on $q_z$ of the structure function.

The next relevant term in the Landau expansion is of the fourth order over $\delta\rho$, it can be written as
 \begin{eqnarray}
  {\cal F}_{(4)}
  =\int \frac{d^3 q_1\, d^3 q_2 d^3 q_3 d^3 q_4}{(2\pi)^9}
  \delta(\bm q_1+\bm q_2+\bm q_3 +\bm q_4)
  \nonumber \\
  \frac{\lambda l}{24}
  \delta\rho(\bm q_1)\delta\rho(\bm q_2)
  \delta\rho(\bm q_3 )\delta\rho(\bm q_4).
 \label{freeweak}
 \end{eqnarray}
The contribution (\ref{freeweak}) describes the self-interaction of the short-range density fluctuations. Generally, the factor $\lambda$ depends on the wave vectors $\bm q_1 \div \bm q_4$. Below, for the sake of simplicity (but keeping an essential physics), we restrict ourselves to the case $\lambda=\mathrm{const}$. One more note of caution is in order here. Since the positional order parameter $\delta \rho$ is a scalar quantity, a third order over $\delta \rho$ contribution into the Landau functional is not forbidden by the symmetry. We neglect this third order term in what follows based on the experimentally confirmed large value of the positional correlation length $\xi_{tr} $. Indeed, if the third-order term is not small then the crystallization of the hexatic smectic phase would be a strong first order transition and $\xi_{tr}$ would be on the order of the molecular size.

The average value of the short-range density modulation $\delta\rho$ is zero in both, smectic A and hexatic, phases. By other words, there is no long-range positional order in the smectic layers. Let us stress that we are interested in the mass density with the wave vectors near the circle $q_\perp = q_0$ (besides the standard for all smectics quasi-Bragg peaks at $q_z=2\pi/l,q_\perp=0$, reflecting the density modulation in the direction perpendicular to the layers). Near the smectic A -- hexatic phase transition (both, above and below the transition temperature $T_h$) the short-range density modulation $\delta\rho$ is a strongly fluctuating quantity, a manifestation of the fact is its large correlation length $\xi_{tr}$. Therefore to find, say, the structure function (\ref{structure}) one has to calculate fluctuation corrections to the bare value determined by the second-order term (\ref{freeweak2}). A theoretical framework of the corresponding analysis can be found in our survey \cite{KLM93}, here we adopt the method for the smectics near the smectic A -- hexatic phase transition.

The interaction between the orientational order parameter and the density fluctuations is described by a crossed term in the Landau functional depending on the both fields, $\delta\rho$ and $\Psi$. The main interaction term in the Landau functional can be written as
 \begin{equation}
 {\cal F}_\mathrm{int}=
 -\frac{1}{2q_0^6}\mathrm{Re}
 \int dV\ \Psi  [(\partial_x-i\partial_y)^3 \delta\rho]^2,
 \label{inter}
 \end{equation}
where, as above, distortions of the smectic layers are neglected. Note that the transformation law (\ref{rotate}) explicitly demonstrates rotational invariance of the interaction term (\ref{inter}). The interaction produces, particularly, a hexagonal angular dependence of the density correlations in the hexatic phase, where the mean value of $\Psi$ is non-zero. Note, that fixing the coefficient in Eq. (\ref{inter}) (that is equal to unity) we define the normalization of the orientational order parameter $\Psi$.

 \section{Fluctuation effects}
 \label{sec:fluct}

To find correlation functions of the order parameters $\Psi$ and $\delta\rho$, one has to take into account a self-interaction of the fields and their coupling. The self-interaction of the hexatic order parameter $\Psi$ leads to a universal scaling behavior characterized by a set of critical exponents \cite{CL00,WK74,ST87,PP79}. The self-interaction of the parameter $\delta\rho$ produces effects that can be examined in the framework of the weak crystallization theory, see Ref. \cite{KLM93}. In addition, we should involve into consideration the coupling between the order parameters $\Psi$ and $\delta\rho$ that will be examined in the framework of the perturbation theory. The applicability condition of this approach is weakness of the coupling.

An analysis of the fluctuational effects shows that in the smectic A phase the structure function (\ref{structure}) can be written as
 \begin{equation}
 S(\bm q)= \frac{T}{\Delta + b(q_\perp-q_0)^2},
 \label{gap}
 \end{equation}
where $b$ is the same parameter as in the Landau functional (\ref{freeweak2}). Below, the parameter $\Delta$ will be termed gap. As it follows from the expression (\ref{gap}), the positional (translational) correlation length is $\xi_{tr} = \sqrt{b/\Delta}$. The expression (\ref{gap}) is correct in the smectic A phase, in the hexatic phase the structure function acquires a hexagonal angular dependence. However, in a vicinity of the the smectic A -- hexatic phase transition, we are interested in this work, the angular dependence is weak and therefore it will be ignored in our analysis. Thus, we use the expression (\ref{gap}) both, for the smectic A and the hexatic phases. Use of the expression (\ref{gap}) is justified by the inequality $q_0 \xi_{tr}\gg1$, that is the main applicability condition for our theory.

The gap $\Delta$ possesses an essential temperature dependence. The bare value of the gap is $a$, see Eq. (\ref{freeweak2}), the value is renormalized due to self-interaction of the density fluctuations. The main fluctuation contribution to $\Delta$ is determined by the so-called one-loop term depicted by the following Feynman diagram
 \begin{equation}
 \bullet\quad \quad \feyn{fl flu}
 \label{feyn1}
 \end{equation}
\vspace{.5cm}

\noindent
where the solid line represents the pair correlation function (\ref{structure}) and the bullet represents $\lambda$, see Eq. (\ref{freeweak}). Adding this fluctuation contribution to the bare value of $\Delta$, that is $a$, one finds the self-consistent equation for the gap
 \begin{equation}
 \Delta = a + \frac{Tq_0 \lambda}{4\sqrt{b\Delta}},
 \label{gap2}
 \end{equation}
in a close analogy to the weak crystallization theory of three-dimensional liquids, see details in the survey \cite{KLM93}. Note that the equation (\ref{gap2}) has a solution for both, positive and negative $a$. That implies that the hexatic phase remains metastable even below the crystallization temperature. Note parenthetically that one has to keep in mind this fact discussing equilibration time in the hexatic smectics.

Next, we should take into account the interaction between the orientational order parameter $\Psi$ and the position order parameter $\delta\rho$ (short-range density fluctuations), determined by the term (\ref{inter}) in the Landau functional. The interaction between the orientation and the position order parameters modifies correlation functions of both fluctuating quantities, $\Psi$ and $\delta\rho$. We examine the effect using the perturbation theory, that is we take into account only first corrections to the correlation functions, related to the interaction (\ref{inter}). The applicability condition of the perturbative approach will be formulated later on in this section.

Let us examine contributions to the gap $\Delta$ caused by the interaction. First contributions to $\Delta$ related to the interaction term (\ref{inter}) can be represented by the following Feynman diagrams
 \begin{equation}
 \feyn{g g \quad \quad fl flu}
 \label{feyn2}
 \end{equation}
\vspace{.5cm}
 \begin{equation}
 \feyn{ f g1 g2 f }
 \label{feyn3}
 \end{equation}
where the wavy lines correspond to the pair correlation function of the hexatic order parameter
\begin{equation}
 F(\bm r_1,\bm r_2)=
 \langle \Psi (\bm r_1)\Psi ^\star(\bm r_2)\rangle.
 \label{pair}
 \end{equation}
According to the second order phase transitions theory \cite{CL00,WK74,ST87,PP79}, the Fourier transform of the correlation function (\ref{pair}) has the following self-similar form
 \begin{equation}
 F(\bm q) = \frac{1}{q^{2 - \eta }} f(q\xi _h),
  \label{pair1}
 \end{equation}
where  $\eta$ is the so-called anomalous critical exponent. For the superfluid universality class $\eta \approx 0.02$, see Refs.  \cite{ST87,AN91,PV02}. The scaling function $f$ in Eq. (\ref{pair1}) provides that $F$ depends solely on $|T - T_h|$ ($T_h$ is the smectic A -- hexatic transition temperature) for $q \xi _h \ll 1$ and solely on $q$ in the opposite limit $q \xi _h \gg 1$ , see Refs. \cite{CL00,WK74,ST87,PP79}.

In the smectic A phase the contribution (\ref{feyn2}) is zero. Indeed, the closed loop in the diagram (\ref{feyn2}) corresponds to the single point average $\langle [(\partial_x-i\partial_y)^3 \delta\rho]^2\rangle$, see Eq. (\ref{inter}), the average is zero due to isotropy of the smectic A layers. In the vicinity of the phase transition, we are considering in this paper, the corresponding contribution is also negligible in the hexatic phase. Thus, one should take into account solely the term (\ref{feyn3}) that gives the following contribution to the gap $\Delta$
 \begin{equation}
 \delta\Delta=-\frac{1}{2T} \int \frac{d^3q}{(2\pi)^3}
 F(\bm q) S(\bm k +\bm q),
 \label{gap3}
 \end{equation}
where $\bm k$ is the wave vector of the density fluctuation. Here the wave vector $k$ lies near the ring (cylinder) $k_\perp=q_0$ whereas the wave vector $q$ can be estimated as $q\sim \xi_h^{-1}$. Thus, the inequalities $k\gg q$ and $\Delta \ll b q^2$ are valid. The first inequality is related to the condition that $\xi_h q_0 \gg 1$, which is valid near the phase transition point, and the second one is equivalent to the condition (\ref{corrl}). Using the inequalities, one obtains
 \begin{eqnarray}
 \delta\Delta =
 -\frac{1}{8\pi^2\sqrt{b\Delta}} \int dq_z dq_\perp F(\bm q),
 \label{delta}
 \end{eqnarray}
where at the derivation of (\ref{delta}) we used the expression (\ref{gap}). Thus we end up with the equation
 \begin{equation}
 \Delta = a + \frac{Tq_0 \lambda}{4\sqrt{b\Delta}} +\delta\Delta,
 \label{gap4}
 \end{equation}
instead of Eq. (\ref{gap2}). The equation (\ref{gap4}) determines the temperature dependence of the gap $\Delta$ in the critical region.

In the region $q\xi _h \gg 1$ the pair correlation function $F(\bm q)\propto q^{\eta-2}$ \cite{CL00,WK74,ST87,PP79}. Therefore there is an ultraviolet contribution to the integral $\int dq_z dq_\perp F$ to be included into a redefinition of the factor $\lambda$, entering Eq. (\ref{gap4}). Besides, there is negative critical contribution to the integral $\int dq_z dq_\perp F$ that behaves $\propto |T-T_h|^{\nu\eta}$, as follows from Eq. (\ref{pair1}). We are interested just in this term that produces a singular contribution to the gap $\Delta$. Taking a derivative of the equation (\ref{gap4}) and keeping in mind that $\Delta$ remains finite at the transition point, we find
 \begin{equation}
 \frac{\partial}{\partial T}\Delta
 \propto  |T-T_h|^{\nu\eta-1}.
 \label{gap7}
 \end{equation}
The singularity is integrable due to $\eta,\nu>0$. Therefore the gap $\Delta$ remains finite at the transition point, indeed, that justifies our approach.

Now we can formulate the applicability condition of our perturbation approach. For the purpose we should compare the contribution depicted by the one-loop diagram (\ref{feyn3}) with contributions of higher order, determined by many-loop diagrams. An example of such diagram is illustrated by the figure
 \begin{equation}
 \feyn{ f g1 g2 f   g4 g3 f}
 \label{feyn33}
 \end{equation}
\vspace{.5cm}

\noindent
where a two-loop diagram is presented. Straightforward estimation shows that one can neglect the diagram (\ref{feyn33}) in comparison with the diagram (\ref{feyn3}) if
 \begin{equation}
 \delta\Delta \ll \Delta,
 \label{criterion}
 \end{equation}
where $\delta\Delta$ is determined by Eq. (\ref{delta}). If the inequality (\ref{criterion}) is not satisfied, we may not restrict ourselves to the one-loop approximation and summation of an infinite series of terms is needed. Such analysis is beyond the scope of this work.

Our theory yields additional contributions to the heat capacity related to the positional degree of freedom $\delta \rho$. The leading contribution is associated with the $T$-dependence of the coefficient $a$ in Eq. (\ref{freeweak2}). Namely, we find for the $a$-dependent part of the free energy
 \begin{equation}
 \frac{\partial F_a}{\partial T}
 = \frac{V}{2} \frac{\partial a}{\partial T}
 \langle (\delta\rho)^2 \rangle
 =\frac{\partial a}{\partial T}
 \frac{TVq_0}{4 l\sqrt{b\Delta}} ,
 \label{gap8}
 \end{equation}
where the thickness $l$ of the smectic layer appears due to integration over $q_z$ within the first Brillouin zone. Taking the $T$-derivative of the expression (\ref{gap8}), we find the following critical contribution to the heat capacity
 \begin{equation}
 -T\frac{\partial^2 F_a}{\partial T^2}
 =\frac{V}{8}\frac{\partial a}{\partial T}
 \frac{T^2 q_0}{l b^{1/2} \Delta^{3/2}}
 \frac{\partial\Delta}{\partial T}.
 \label{gap9}
 \end{equation}
In accordance with Eq. (\ref{gap7}), the contribution (\ref{gap9}) diverges near the phase transition with the exponent $1-\nu\eta$, close to unity.

The additional contribution to the heat capacity (\ref{gap9}) is crucial for our approach. The main message of our work (see details in the next section) is to claim that a sum of two critical contributions to the heat capacity, with the ``small'' exponent $\alpha$ and with the ``large'' exponent $1-\nu\eta$, enables one to describe quantitatively the known experimental data.

\section{Comparison with experiment}
\label{sec:exp}

Unfortunately in the critical region for the smectic $A$ - hexatic phase transition, the X-ray scattering experimental data are very scarce (only a few experimental points). Certainly in such a situation it is useless to fit the theory to estimate at least 3 unknown parameters entering the theory equation for the gap. On the contrary, calorimetric data in the critical region are quite informative. Thus in this work we focus on the measurable specific heat behavior and take a pragmatic approach to understand calorimetric features of the smectic $A$
-- hexatic smectic phase transition, and provide relationships between different physical properties.

Let us turn to comparing our theory with the heat capacity experimental data. Known from the literature experimental calorimetric data for 65OBC liquid crystalline material \cite{VL83,HV83,PN85,GS91,SG92,GS93,HK97,RD05,MP13} manifest that the smectic A -- hexatic transition is a second order phase transition with a strong, nearly symmetric singularity of the heat capacity. The value of the critical exponent $\alpha$ is extracted in the works from fitting the experimental data to a single power law dependence in the temperature range $|T-T_h| > 0.1 K$. For 65OBC the effective critical exponent is $\alpha \approx 0.64$, and the critical amplitude ratio is $A^+/A^- \approx 0.84$. Note that the effective critical exponent $\alpha$ depends on the material. Results presented in the work \cite{PN85} give the values ranging from $0.48$ to $0.67$ for eight different substances. Such dispersion signals about at least some problems of the approach.

Our theory states that the singular part of the heat capacity in the intermediate critical region is a sum of the two terms: the first term with the standard small critical exponent $\alpha$ characteristic of the theory for the two-component order parameter universality class and the second term with the exponent $1-\nu\eta\approx 0.985$, see Eq. (\ref{gap9}). The second contribution originates from the interaction of the order parameters $\Psi$ and $\delta\rho$. Thus to fit the experimental data we utilize the following expression for the heat capacity
 \begin{eqnarray}
 C=\frac{p_1}{|x|^{-0.013}}
 +\frac{p_3}{|x|}+p_5,
 \quad \mathrm{if}\ x<0,
 \label{new} \\
 C=\frac{p_2}{x^{-0.013}}
 +\frac{p_4}{x}+p_5,
 \quad \mathrm{if}\ x>0,
 \nonumber
 \end{eqnarray}
where $x=(T-T_h)/T_h$ is the reduced temperature, and we borrowed the value of the exponent $\alpha = - 0.013$ from the standard data for the two-component order parameter in three dimensions presented in \cite{ST87,AN91,PV02}. The obtained values giving the best fitting are $T_h=341.11 K$, $p_1=-48.09599$, $p_2=-48.19495$, $p_3=0.0008$, $p_4=0.00064$ and $p_5=91.60242$. Although the found dimensionless parameters $p_3$ and $p_4$ are much smaller than $p_1$ and $p_2$, the former ones are meaningful and anyway larger than the numeric accuracy of the fitting procedure. Physically, the small values of the parameters $p_3$ and $p_4$ indicate (consistently with our perturbation theory assumptions) that the coupling term (\ref{inter}) is small.

\begin{figure}
\begin{center}
\includegraphics[height=2.5in]{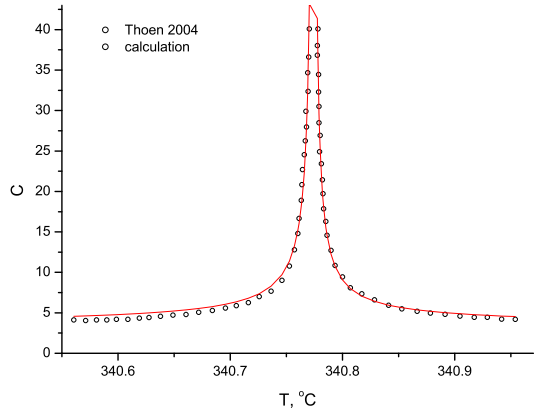}
\end{center}
\caption{Calorimetric experimental data \cite{RD05} for 65OBC (open circles) versus our calculation (solid line)}
\label{f2}
\end{figure}

Figure \ref{f2} confronts results of the computations of the specific heat in our model, see Eqs. (\ref{gap9}) and (\ref{new}), and the experimental data \cite{RD05} (which almost coincide, up to a regular shift of the sample-dependent transition temperature, with the data presented in Refs. \cite{VL83} and \cite{HK97}).  We see a reasonable agreement of the experimental data and our theory (valid in the temperature range about $0.6\, -\, 1\, K$ around $T_h$). The exponent $\alpha $ found in the works \cite{VL83}, \cite{HK97}, \cite{RD05} is based on the single power-law fitting of the experimental data in the temperature range $0.005 < |T-T_h|/T_h < 1.5$. In this more broad temperature region we expect a sort of crossover behavior which can be fitted by a single (but not universal!) power law. Understanding all its limitations, we are confident that our model captures the essential features of the smectic A -- hexatic phase transition in the intermediate critical region (with two strongly fluctuating and coupled order parameters).

\section{Conclusion}
\label{sec:con}

The main advance of our work is establishing for the first time the model enabling to match quantitatively theory and calorimetric data for the smectic A -- hexatic phase transition, explaining, particularly, the effective ``large'' value of the heat capacity critical exponent. To avoid a confusion, let us stress that we are not fighting with the well established and widely accepted conception of the second order phase transition universality. There are no doubts that sufficiently close to the smectic A -- hexatic transition temperature, the universal behavior of the heat capacity has to be observed. However, we claim that the temperature region where the standard universality holds in a narrow vicinity of the transition point. Outside the region the critical behavior does not correspond to the standard universality, its description needs some additional ingredients, presented in our paper.

Our model involves an interaction between the strongly fluctuating orientational (hexatic) and positional degrees of freedom, $\Psi$ and $\delta\rho$, that offers a new non-trivial scenario for the temperature dependencies of the specific heat and X-ray scattering patterns near the smectic A -- hexatic phase transition. Our approach allows one to reconcile well established phase transition universality conception and the calorimetric experimental data \cite{VL83,HV83,PN85,GS91,SG92,GS93,HK97,RD05,MP13} for the smectic A -- hexatic smectic phase transition. Thus, we claim, that we solved the old riddle related to the ``large'' value of the heat capacity exponent for the phase transition. Our theoretical model lays firm foundations for the further studies for biologists, experimental and theoretical physicists.

Our results, reported in the paper, concern the critical region (region of strong fluctuations of the hexatic order parameter $\Psi$) near the transition point. However, our model can be applied to the hexatic phase (outside the critical region) as well. In the region one has to take into account fluctuations of both, $\Psi$ and $\delta\rho$. Certain features of the fluctuations in the hexatic phase are different from those in the critical region due to non-zero mean value of the order parameter $\Psi$. A description of the region is a subject for future investigations.

One should keep in mind that our scheme implies weakness of the interaction between the orientation and the position degrees of freedom. That is why we restrict ourselves to first corrections over the interaction. An intriguing question about the system behavior when the coupling is not small is beyond the scope of our consideration in this publication, it will be analyzed elsewhere.

\acknowledgements

Our work was funded by Russian Science Foundation (grant 14-12-00475). We acknowledge stimulating discussions with B.I.Ostrovskii, I.A.Vartanyants, and I.A.Zaluzhnyy, inspired this work. This paper has benefited also from
exchanges with many colleagues. We thank in particular I.Kolokolov and L.Schur.


\begin{thebibliography}{99}

 \bibitem{BL78}
 R.J.Birgeneau, J.D.Litster,
 J. Phys. (Paris) {\bf 39}, 1399 (1978).

 \bibitem{BN81}
 R.Bruinsma, D.R.Nelson,
 Phys. Rev. B {\bf 23}, 402 (1981).

 \bibitem{PM81}
 R.Pindak, D.E.Moncton, S.C.Davey, J.W.Goodby,
 Phys. Rev. Lett. {\bf 46}, 1135 (1981).

 \bibitem{BA86}
 J.D.Brock, A.Aharony, R.J.Birgeneau, K.W.Evans-Lutterodt, J.D.Litster, P.M.Horn, G.B.Stephenson, A.R.Tajbakhsh,
 Phys. Rev. Lett. {\bf 57}, 98 (1986).

 \bibitem{JO03}
 W.H. de Jeu, B.I.Ostrovskii, and A.N.Shalaginov,
 Rev. Mod. Phys. {\bf 75}, 181 (2003).

 \bibitem{ZK15}
 I.A. Zaluzhnyy, R.P.Kurta, E.A.Sulyanova, O.Y.Gorobtsov, A.G.Shabalin, A.V.Zozulya,
 A.P. Menushenkov, M.Sprung, B.I.Ostrovskii, I.A.Vartanyants,
 Phys. Rev. E {\bf 91}, 042506 (2015).

 \bibitem{NT00}
 J.F. Nagle, S. Tristram-Nagle,
 Biochim. Bophys. Acta {\bf 10}, 159 (2000).

 \bibitem{MR05}
 S.J.Marrink, H.J.Risselada, A.E.Mark,
 Chemistry and Physics of Lipids {\bf 135}, 223 (2005).

 \bibitem{PV15}
 O.F.Petrov, M/M.Vasiliev, O.S.Vaulina, K.B.Stasenko, E.K.Vasilieva, E.A.Lisin, Y.Tun, V.E.Fortov,
 Europhys. Lett. {\bf 111}, 45002 (2015).

\bibitem{VL83}
 J.M.Viner, D.Lamey, C.C.Huang, R.Pindak, J.W.Goodby,
 Phys. Rev. A {\bf 28}, 2433 (1983).

  \bibitem{HV83}
  C.C. Huang, J.M. Viner,R. Pindak, J.W. Goodby,
  Phys. Rev. Lett. {\bf 46}, 1289 (1981).

 \bibitem{PN85}
 T. Pitchford, G. Nounesis, S. Dumrongrattana, J.M. Viner, C.C. Huang, J. W. Goodby,
 Phys. Rev. A {\bf 32}, 1938 (1985).

 \bibitem{GS91}
 R.Geer, T.Stoebe, C.C.Huang, R.Pindak, G.Srajer, J.W.Goodby, M.Cheng, J.T.Ho, S.W.Hui,
 Phys. Rev. Lett. {\bf 66}, 1322 (1991).

 \bibitem{SG92}
 T.Stoebe, R.Geer, C.C.Huang, J.W.Goodby,
 Phys. Rev. Lett. {\bf 69}, 2090 (1992).

 \bibitem{GS93}
 R.Geer, T.Stoebe, C.C.Huang,
 Phys. Rev. E {\bf 48}, 408 (1993).

 \bibitem{HK97}
 H. Haga, Z. Kutnjak, G.S. Iannacchione, S. Qian, D. Finotello, C.W. Garland,
 Phys. Rev. E {\bf 56}, 1808 (1997).

 \bibitem{RD05}
 B. Van Roie, K.Denolf, G.Pitsi, J.Thoen,
 Eur. Phys. J. E {\bf 16}, 361 (2005).

 \bibitem{MP13}
 F.Mercuri, S.Paolini, M.Marinelli, R.Pizzoferrato, U.Zammit,
 J. Chem. Phys. {\bf 138}, 074903 (2013).

 \bibitem{CL00}
 P.M.Chaikin, T.C.Lubensky,
 Principles of condensed matter physics, Cambridge University Press, Cambridge, 2000.

 \bibitem{GP93}
 P.G. de Gennes and J. Prost,
 The Physics of Liquid Crystals, Claredon Press, Oxford, 1993.

 \bibitem{GKL91}
 E.V.Gurovich, E.I.Kats, V.V.Lebedev,
 ZhETF {\bf 100}, 855 (1991) [Sov. Phys. JETP {\bf 73}, 473 (1991)].

 \bibitem{LL80}
 L.D.Landau, E.M.Lifshitz, Course of Theoretical Physics,
 Statistical Physics, Part 1, Pergamon Press, New York, 1980.

 \bibitem{WK74}
 K.G.Wilson and J.Kogut, Physics Reports {\bf 12}, 75 (1974).

 \bibitem{ST87}
 H.E.Stanley, Introduction to phase transitions and critical phenomena,
 Oxford University Press, New York, 1987.

 \bibitem{AN91}
 M.A.Anisimov, Critical phenomena in liquids and liquid crystals,
 Gordon and Breach, Philadelphia, 1991.

 \bibitem{PV02}
 A.Pelissetto, E.Vicari,
 Physics Reports {\bf 368}, 549 (2002).

 \bibitem{GZ80}
 J.C. Le Guillou, J.Zinn-Justin,
 Phys. Rev. B {\bf 21}, 3976 (1980).

 \bibitem{AB86}
 A.Aharony, R.J.Birgeneau, J.D.Brock, J.D.Litster,
 Phys. Rev. Lett. {\bf 57}, 1012 (1986).

 \bibitem{AV85}
 M.A.Anisimov, V.P.Voronov, A.O.Kul'kov, F.Kholmurodov,
 Pis'ma v ZhETF {\bf 41}, 248 (1985) [JETP Lett. {\bf 41}, 302 (1985)].

 \bibitem{RH82}
 C.Rosenblat, J.T.Ho,
 Phys. Rev. A {\bf 26}, 2293 (1982).

 \bibitem{BB89}
 J.D.Brock, R.J.Birgeneau, D.Litster, A.Aharony,
 Contemp. Phys. {\bf 3}, 321 (1989).

 \bibitem{GC94}
 E.Gorecka, Li Chen, O.Lavrentovich, W.Pyzuk,
 Europhys. Lett. {\bf 27}, 507 (1994).

 \bibitem{BK76}
 T.W.Burkhadt, H.J.F. Knops, M. de Vijs,
 J. of Phys. A {\bf 9}, L-179 (1976).

 \bibitem{KLM93}
 E.I. Kats, V.V. Lebedev, A.R. Muratov,
 Physics Reports {\bf 228}, 1 (1993).

 \bibitem{PP79}
 A.Z. Patashinskii, V.L. Pokrovskii,
 Fluctuation Theory of Phase Transitions, Pergamon Press, New York, 1979.

 \end{thebibliography}
\end{document}